\documentclass{emulateapj}

\usepackage{amssymb}

\def\ts     {\thinspace}
\def\kms    {\ts km\ts s$^{-1}$}
\def\etal   {{\rm et\ts al.}}

\def\cco    {{\rm CO}($J$=3$\to$2)}

\def\gco    {{\rm CO}($J$=7$\to$6)}

\def\ahcn    {{\rm HCN}($J$=1$\to$0)}

\def\dhcn    {{\rm HCN}($J$=4$\to$3)}

\def\ahco    {{\rm HCO$^+$}($J$=1$\to$0)}

\def\dhco    {{\rm HCO$^+$}($J$=4$\to$3)}

\def\ccn     {{\rm CN}($N$=3$\to$2)}

\slugcomment{draft version \today, accepted for publication in the Astrophysical Journal}

\shorttitle{Dense Molecular Gas Excitation in the Cloverleaf}
\shortauthors{Riechers et al.}

\begin{document}

\title{
Dense Molecular Gas Excitation at High Redshift:\\
Detection of HCO$^+$($J$=4$\to$3) Emission in the Cloverleaf Quasar
}

\author{Dominik A.\ Riechers\altaffilmark{1,2,7}, Fabian Walter\altaffilmark{2}, Christopher L.\ Carilli\altaffilmark{3}, \\ Pierre Cox\altaffilmark{4}, Axel Wei\ss\altaffilmark{5}, Frank Bertoldi\altaffilmark{6}, and Karl M.\ Menten\altaffilmark{5}}

\altaffiltext{1}{Astronomy Department, California Institute of
  Technology, MC 249-17, 1200 East California Boulevard, Pasadena, CA
  91125, USA; dr@caltech.edu}

\altaffiltext{2}{Max-Planck-Institut f\"ur Astronomie, K\"onigstuhl 17, D-69117 Heidelberg, Germany}

\altaffiltext{3}{National Radio Astronomy Observatory, PO Box O, Socorro, NM 87801, USA}

\altaffiltext{4}{Institut de RadioAstronomie Millim\'etrique, 300 Rue
  de la Piscine, Domaine Universitaire, 38406 Saint Martin d'H\'eres,
  France}

\altaffiltext{5}{Max-Planck-Institut f\"ur Radioastronomie, Auf dem H\"ugel 69, Bonn, D-53121, Germany}

\altaffiltext{6}{Argelander-Institut f\"ur Astronomie, Universit\"at
  Bonn, Auf dem H\"ugel 71, Bonn, D-53121, Germany}

\altaffiltext{7}{Hubble Fellow}


\begin{abstract}

We report the detection of \dhco\ emission in the Cloverleaf Quasar at
$z$=2.56, using the IRAM Plateau de Bure Interferometer. HCO$^+$
emission is a star formation indicator similar to HCN, tracing dense
molecular hydrogen gas ($n({\rm H_2}) \simeq 10^5\,$cm$^{-3}$) within
star-forming molecular clouds.  We derive a lensing-corrected \dhco\
line luminosity of $L'_{\rm HCO^+(4-3)} = (1.6 \pm 0.3) \times
10^{9}\,(\mu_{\rm L}/11)^{-1}\,$K\,\kms\,pc$^2$, which corresponds to
only 48\% of the \ahco\ luminosity, and $\lesssim$4\% of the \cco\
luminosity. The HCO$^+$ excitation thus is clearly subthermal in the
$J$=4$\to$3 transition. Modeling of the HCO$^+$ line radiative
transfer suggests that the HCO$^+$ emission emerges from a region with
physical properties comparable to that exhibiting the CO line
emission, but 2$\times$ higher gas density. This suggests that both
HCO$^+$ and CO lines trace the warm, dense molecular gas where star
formation actively takes place. The HCO$^+$ lines have only $\sim$2/3
the width of the CO lines, which may suggest that the densest gas is
more spatially concentrated.  In contrast to the $z$=3.91 quasar
APM\,08279+5255, the dense gas excitation in the Cloverleaf is
consistent with being purely collisional, rather than being enhanced
by radiative processes. Thus, the physical properties of the dense gas
component in the Cloverleaf are consistent with those in the nuclei of
nearby starburst galaxies.  This suggests that the conditions in the
dense, star-forming gas in active galactic nucleus-starburst systems
at early cosmic times like the Cloverleaf are primarily affected by
the starburst itself, rather than the central active black hole.

\end{abstract}

\keywords{galaxies: active --- galaxies: starburst --- 
galaxies: formation --- galaxies: high-redshift --- cosmology: observations 
--- radio lines: galaxies}

\section{Introduction}

Investigating the dense molecular interstellar medium (ISM) in distant
galaxies is of fundamental importance for our general picture of
galaxy formation and evolution in the early universe, as it is found
in the regions where active star formation occurs. Due to the fact
that CO exhibits the brightest emission lines of all molecules, it is
a good tracer for molecular clouds and the diffuse, gaseous ISM; i.e.,
the total amount of potential fuel for star formation (see Solomon \&
Vanden Bout \citeyear{sv05} for a review). However, the low density
required to excite CO ($>300$\,cm$^{-3}$) also means that it is not a
specific tracer of the molecular cloud cores where star formation
actively takes place. In contrast, recent studies have shown that high
dipole moment molecules like HCO$^+$ and HCN are substantially better
tracers of cloud cores. This is due to the fact that such molecules
trace much denser gas ($>10^{5}$\,cm$^{-3}$) than CO emission (e.g.,
Gao \& Solomon \citeyear{gs04a}, \citeyear{gs04b}).

As they only trace the densest regions of the ISM (and are less
abundant), emission from rotational transitions of dense gas tracers
is typically by at least an order of magnitude fainter than emission
from CO lines. Thus, studies of the dense ISM in high redshift
galaxies are currently focused on only few, bright
targets. Consequently, high-$z$ HCN and HCO$^+$ emission was first
detected in the Cloverleaf quasar ($z$=2.56), the brightest known CO
line emitter at the time (Solomon et al.\
\citeyear{sol03}; Riechers et al.\ \citeyear{rie06}). Both molecules 
were detected in the ground-state $J$=1$\to$0 transitions, which trace
the full amount of dense gas. Despite a number of physical and
chemical processes that affect HCN and HCO$^+$ in different ways,
their respective $J$=1$\to$0 lines show comparable strengths,
indicating that the emission is likely optically thick.

To understand the physical conditions in the dense, star-forming
molecular gas in more detail, it is necessary to study multiple
rotational lines from dense gas tracers. Previous searches for \dhco\
and \dhcn\ line emission in the Cloverleaf quasar were unsuccessful
(Wilner et al.\ \citeyear{wil95}; Solomon et al.\ \citeyear{sol03};
see also Barvainis et al.\ \citeyear{bar97}). The limits indicate a
comparatively low excitation of HCN, but are not constraining for
HCO$^+$, given the measured $J$=1$\to$0 line luminosities.

\begin{figure*}
\vspace{-2mm}
\epsscale{1.15}
\plotone{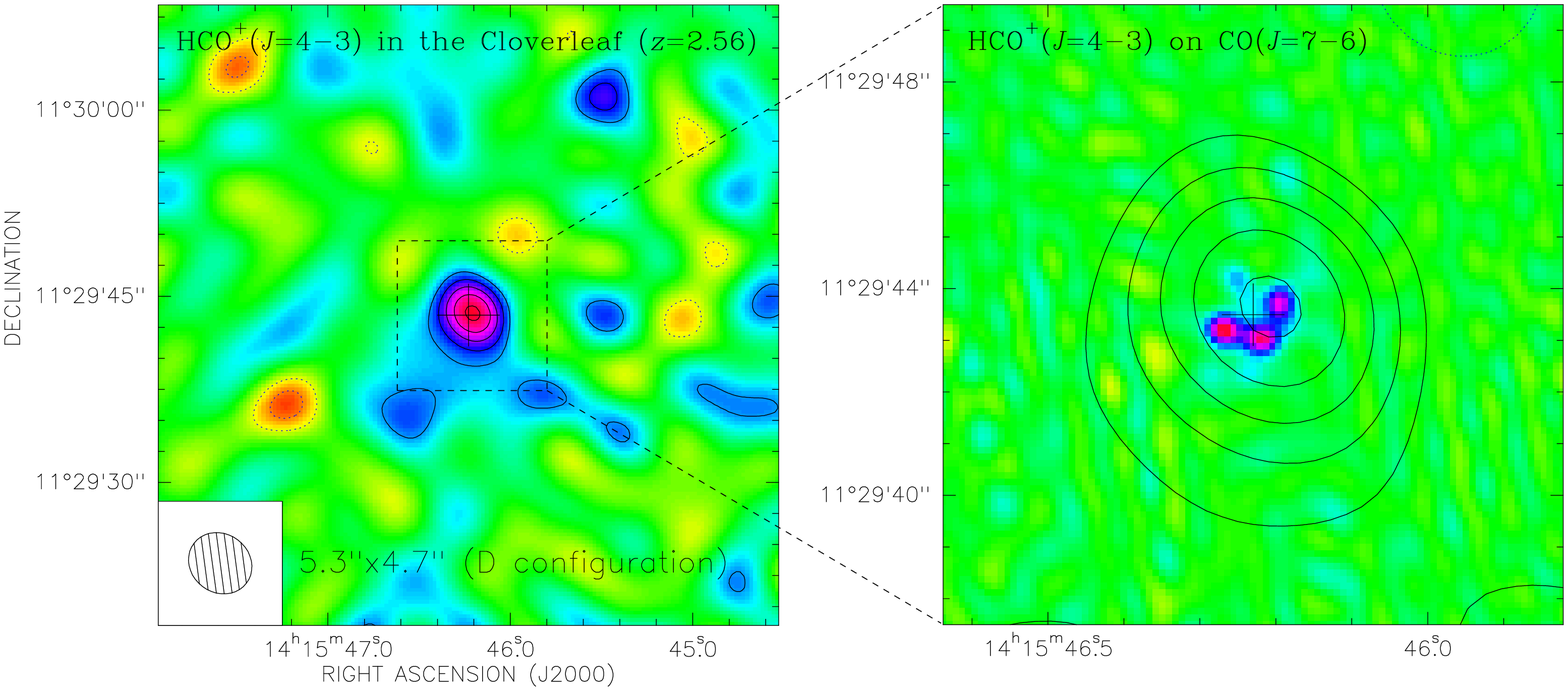}
\vspace{-2mm}

\caption{Velocity-integrated PdBI map of \dhco\ line emission over 
284\,\kms\ (approximately the line FWHM) toward the Cloverleaf. At a
resolution of 5.3$''$$\times$4.7$''$ (as indicated in the bottom
left), the emission remains unresolved (maximum lens image
separation:\ $\sim$1.4$''$). The cross indicates the center of the
\gco\ emission (color scale in the {\em right} panel; Alloin \etal\
\citeyear{all97}). Contours are shown in steps of
1$\sigma$=0.31\,mJy\,beam$^{-1}$, starting at $\pm$2$\sigma$.
\label{f1}}
%
\end{figure*}

In this Letter, we report the detection of \dhco\ emission toward the
Cloverleaf quasar ($z$=2.56), using the IRAM Plateau de Bure
Interferometer (PdBI). This enables us to investigate the dense
molecular gas excitation in a high-$z$ galaxy, constraining the
physical conditions for star formation out to the early universe. We
use a concordance, flat $\Lambda$CDM cosmology throughout, with
$H_0$=71\,\kms\,Mpc$^{-1}$, $\Omega_{\rm M}$=0.27, and
$\Omega_{\Lambda}$=0.73 (Spergel \etal\
\citeyear{spe03}, \citeyear{spe07}).

\section{Observations}

We observed the \dhco\ transition line ($\nu_{\rm rest} =
356.734288\,$GHz) toward H1413+117 (the Cloverleaf quasar), using the
PdBI. At the target redshift of $z$=2.55784 (e.g., Barvainis et al.\
\citeyear{bar94}; Wei\ss\ \etal\ \citeyear{wei03}), the line is shifted
to 100.267\,GHz (2.99\,mm).  A total bandwidth of 580\,MHz
($\sim$1700\,\kms ) was used to cover the \dhco\ line and the
underlying 3\,mm (rest-frame 840\,$\mu$m) continuum emission.
Observations were carried out under acceptable to good 3\,mm weather
conditions during 10\,tracks in D configuration between 2006 May 12
and August 20. The total integration time amounts to 70.2\,hr using 4,
5, or 6 antennas, resulting in 10.8\,hr 6 antenna-equivalent on-source
time after discarding unusable visibility data. The nearby sources
B1354+195 and B1502+106 (distance to the Cloverleaf: $9.0^\circ$ and
$12.0^\circ$) were observed every 20 minutes for pointing, secondary
amplitude and phase calibrations.  For primary flux and bandpass
calibration, several nearby calibrators (MWC349, CRL618, 3C84, 3C273,
3C279, 3C345, and 3C454.3) were observed during all runs, leading to a
calibration that is accurate within 10--15\%.

For data reduction and analysis, the IRAM GILDAS package was used.
All data were mapped using the CLEAN algorithm and `natural'
weighting; this results in a synthesized beam of
5.3\,$''$$\times$4.7\,$''$ ($\sim$40\,kpc at $z$ = 2.56).  The final
rms is 0.31\,mJy beam$^{-1}$ over 95\,MHz (corresponding to
284\,\kms), and 0.79\,mJy beam$^{-1}$ over 15\,MHz (45\,\kms).

\section{Results}

We have detected \dhco\ line emission toward the Cloverleaf quasar
($z$=2.56) at 6$\sigma$ significance (Fig.~\ref{f1}). From Gaussian
fitting to the line profile, we obtain a \dhco\ line peak strength of
1.77$\pm$0.35\,mJy at a FWHM of 288$\pm$79\,\kms, on top of
0.29$\pm$0.14\,mJy continuum emission (consistent with
0.3$\pm$0.1\,mJy as measured at 93\,GHz and the dust spectral energy
distribution; Henkel et al.\ \citeyear{hen10}; Wei\ss\ et al.\
\citeyear{wei03}). This corresponds to a velocity-integrated emission
line strength of 0.54$\pm$0.09\,Jy\,\kms, and a line luminosity of
$L'_{\rm HCO^+(4-3)} = (1.6 \pm 0.3) \times 10^{9}\,(\mu_{\rm
L}/11)^{-1}\,$K\,\kms\,pc$^2$ (where $\mu_{\rm L}$=11 is the lensing
magnification factor; Venturini \& Solomon \citeyear{vs03}), i.e.,
only $\lesssim$4\% of the \cco\ luminosity, and $\sim$35\% of the
\ccn\ luminosity (Wei\ss\ \etal\ \citeyear{wei03}; Riechers et al.\ 
\citeyear{rie07}).

The line FWHM is only $\sim$70\% of that of the \cco\ line (Wei\ss\ et
al.\ \citeyear{wei03}). With this new constraint, we re-visited the
\ahco\ line data by Riechers et al.\ (\citeyear{rie06}; taken with 
substantially narrower bandwidth), which already showed some evidence
for a narrower line than CO. The \ahco\ data can be fitted well with a
(beam-corrected) line peak strength of 0.249$\pm$0.028\,mJy at a FWHM
of 263$\pm$52\,\kms.\footnote{This width is also consistent with that
of the \ahcn\ line (Solomon et al.\ \citeyear{sol03}).} This corresponds to a
velocity-integrated emission line strength of
0.069$\pm$0.008\,Jy\,\kms, and a line luminosity of $L'_{\rm
HCO^+(1-0)} = (3.3 \pm 0.3) \times 10^{9}\,(\mu_{\rm
L}/11)^{-1}\,$K\,\kms\,pc$^2$, which is consistent with the previously
derived value (Riechers et al.\ \citeyear{rie06}) within the
errors. More importantly, this corresponds to a HCO$^+$
$J$=4$\to$3/1$\to$0 line brightness temperature ratio of
$r_{41}$=0.48$\pm$0.11, i.e., the \dhco\ line is clearly subthermally
excited.\footnote{Or optically thin, but see Sect.~4.} This is
consistent with the previous limit of $r_{41}$$<$4 (Wilner et al.\
\citeyear{wil95}; Riechers et al.\ \citeyear{rie06}).  We also set a
3$\sigma$ lower limit of $r_{4}({\rm HCO^+/HCN})$$>$0.59 on the
HCO$^+$/HCN $J$=4$\to$3 line ratio, consistent with the $J$=1$\to$0
line ratio of $\sim$0.8 within the errors (Riechers et al.\
\citeyear{rie06}; HCN $J$=4$\to$3 limit from Solomon et al.\
\citeyear{sol03}).

\begin{figure}
\epsscale{1.15}
\plotone{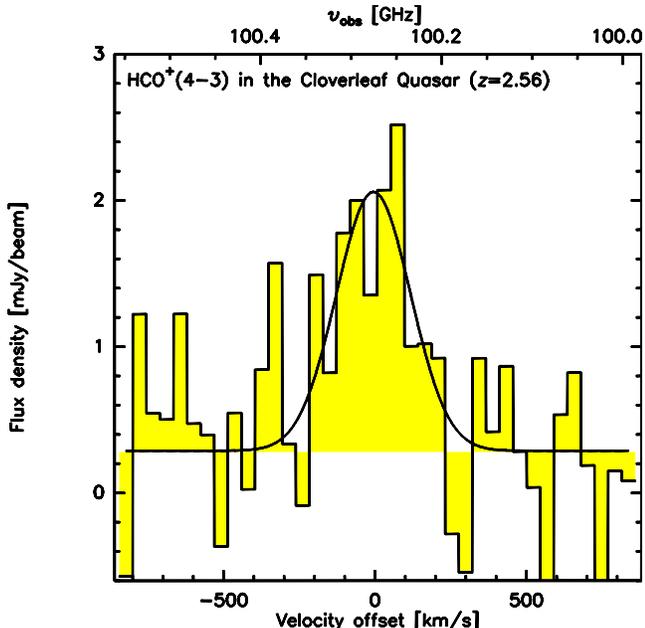}
\vspace{-2mm}

\caption{{\em Left}:\ PdBI \dhco\ spectrum of the Cloverleaf at 15\,MHz (45\,\kms ) 
resolution (histogram), along with a Gaussian fit to the line and
continuum emission (black curve). The line FWHM is 288$\pm$79\,\kms.
The velocity scale is relative to the source's redshift of $z$=2.55784
(e.g., Barvainis et al.\ \citeyear{bar94}; Wei\ss\ et al.\
\citeyear{wei03}).
\label{f2}}
%
\end{figure}

\section{HCO$^+$ and CO Line Excitation Modeling}

Based on the HCO$^+$ excitation ladder of the Cloverleaf, we can
constrain the line radiative transfer of the dense molecular gas
component through Large Velocity Gradient (LVG) models, treating the
gas kinetic temperature and density as free parameters. To maximize
the available observational constraints on these parameters, we here
simultaneously model the HCO$^+$ and CO excitation, and also require
that our solutions are consistent with the dust spectral energy
distribution of the Cloverleaf (Wei\ss\ et al.\ \citeyear{wei03}).

\begin{figure*}
\vspace{-2mm}
\epsscale{1.0}
\plotone{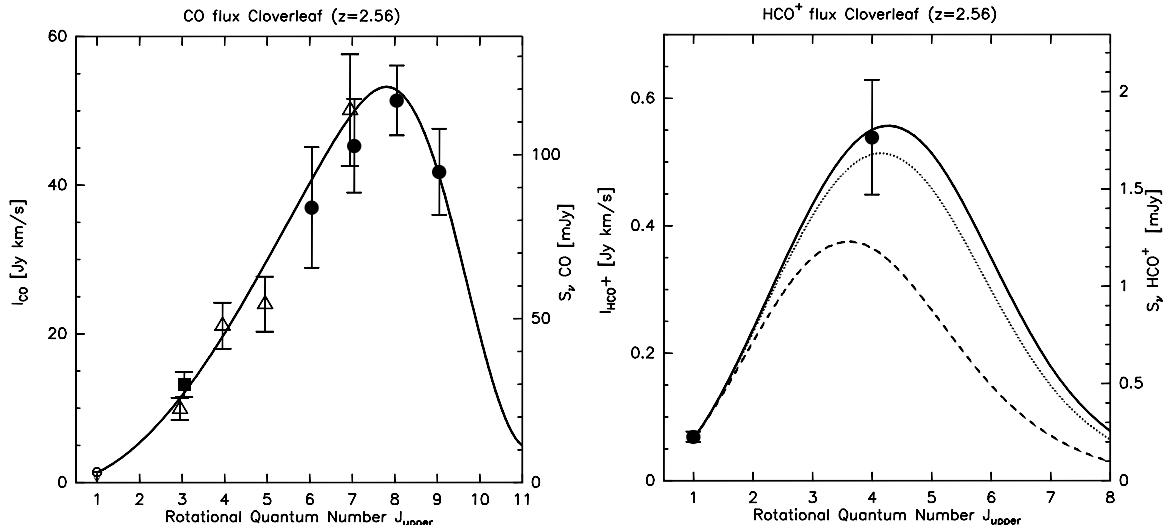}
\vspace{-2mm}

\caption{CO ({\em left}) and HCO$^+$ ({\em right}) excitation ladders 
(points) and LVG models (lines) for the Cloverleaf.  {\em Left}:\ The
CO data (circle:\ Tsuboi et al.\ \citeyear{tsu99}; triangles:\
Barvainis et al.\ \citeyear{bar97}; filled square:\ Wei\ss\ et al.\
\citeyear{wei03}; filled circles:\ Bradford et al.\ \citeyear{bra09})
are fit well by a gas component with a kinetic temperature of $T_{\rm
kin}$=50\,K and a gas density of $\rho_{\rm
gas}$=10$^{4.5}$\,cm$^{-3}$. {\em Right}:\ The HCO$^+$ data are fit
well by a dense gas component with $T_{\rm kin}$=50\,K and $\rho_{\rm
gas}$=10$^{4.8}$\,cm$^{-3}$ (i.e., 2$\times$$\rho_{\rm gas}$ of CO;
solid line). For comparison, the dashed line shows a model with the
same $\rho_{\rm gas}$ as in the CO model. The dotted line shows a
model with the same $\rho_{\rm gas}$ as in the CO model, and a
10$^{0.5}$$\times$ higher HCO$^+$ abundance as in the other
models. All models are scaled to the same \ahco\ flux.
\label{f3}}
%
\end{figure*}

Our models use the HCO$^+$ and CO collision rates from Flower
(\citeyear{flo99}, \citeyear{flo01}). We adopt a HCO$^+$ abundance per
velocity gradient of [HCO$^+$]/(${\rm d}v/{\rm d}r) = 1 \times
10^{-9}\,{\rm pc}\,$(\kms)$^{-1}$, and [HCO$^+$/CO]=10$^{-4}$ (e.g.,
Wang et al.\ \citeyear{wan04}). The HCO$^+$ data are fit well by a
spherical, single-component model with a kinetic temperature of
$T_{\rm kin}$=50\,K, a gas density of $n_{\rm
gas}$=10$^{4.8}$\,cm$^{-3}$, and a CO disk filling factor of 22\%
($r_{\rm CO}$=785\,pc, assuming $\mu_{\rm L}$=11; {\em right} panel in
Fig.~\ref{f3}). In this model, the HCO$^+$ $J$=1$\to$0 and 4$\to$3
lines have optical depths of $\tau_{1-0}$=3.1 and $\tau_{4-3}$=25.4
and excitation temperatures of $T_{\rm ex}^{1-0}$=36.7\,K and $T_{\rm
ex}^{4-3}$=24.4\,K. This suggests that the emission in both
transitions is optically thick, and that the
\dhco\ line, indeed, is subthermally excited.

The (literature) CO data are fit well by a spherical, single-component
model with a kinetic temperature of $T_{\rm kin}$=50\,K, a gas density
of $n_{\rm gas}$=10$^{4.5}$\,cm$^{-3}$, and a $\sim$5.5$\times$ higher
surface filling factor than HCO$^+$ ({\em left} panel in
Fig.~\ref{f3}). These parameters are compatible with the CO excitation
analysis of Bradford et al.\ (\citeyear{bra09}). Our analysis thus
suggests that the CO and HCO$^+$ excitation in the Cloverleaf can be
modeled simultaneously, with the same kinetic temperatures and only a
factor of 2 difference in gas densities.

To explore the remaining uncertainties, we also attempted to fit the
HCO$^+$ data with $T_{\rm kin}$ and $n_{\rm gas}$ fixed to those of
the CO model, but varying the relative molecular abundance. We find an
acceptable solution when increasing the relative HCO$^+$ abundance to
[HCO$^+$]/(${\rm d}v/{\rm d}r) = 1 \times 10^{-8.5}\,{\rm
pc}\,$(\kms)$^{-1}$, i.e., [HCO$^+$/CO]=10$^{-3.5}$, and the HCO$^+$
surface filling factor to 24\% (see dashed and dotted lines in
Fig.~\ref{f3}).

Overall, the models suggest a relatively high median gas density in
this galaxy, and that the CO and the HCO$^+$ emission likely trace the
same warm, dense molecular ISM phase, with HCO$^+$ tracing the densest
$\sim$15--20\% of the gas. Given that the HCO$^+$ lines have only
$\sim$2/3 of the width of the CO lines, this may suggest that the
emission region with the densest gas is more spatially concentrated
than the overall CO emission, such as, e.g., in a nuclear
starburst. Higher spatial resolution HCO$^+$ observations are required
to confirm this scenario, and to investigate potential differential
lensing effects.

\section{Discussion}

We have detected bright, but subthermally excited \dhco\ emission
toward the Cloverleaf quasar at $z$=2.56. Based on excitation
modeling, we find that the warm, dense gas traced by HCO$^+$ appears
to be associated with the warm gas phase traced by the CO lines,
picking out its densest regions.

The HCO$^+$ excitation in the Cloverleaf is consistent with that seen
in the starburst nucleus of NGC\,253, with HCO$^+$ $J$=4$\to$3/1$\to$0
ratios of $r_{41}$=0.48$\pm$0.11 and 0.53 (Knudsen et al.\
\citeyear{knu07}), respectively. Intriguingly, the Cloverleaf and the
nucleus of NGC\,253 also have comparable HCO$^+$/HCN $J$=1$\to$0 line
ratios of $\sim$0.8. The $r_{41}$ in the Cloverleaf is higher than
that in the infrared-luminous galaxies NGC\,6240 (0.21$\pm$0.06) and
Arp\,220 (0.33$\pm$0.12; Greve et al.\ \citeyear{gre09}). However,
this is likely due to the fact that the line ratios are averaged over
virtually the entire molecular line emission regions, rather than just
the nuclei. As shown by Iono et al.\ (\citeyear{ion07}; their
Fig.~10), the $r_{41}$ in NGC\,6240 scatters up to values of $\sim$0.6
within the HCO$^+$-emitting region, indicating that the ratio is
comparable to the Cloverleaf in the densest regions. Thus, it seems
plausible that the dense gas excitation in the Cloverleaf is
comparable to what is found in the nuclear regions of nearby starburst
galaxies and luminous infrared galaxies.

The relatively high median gas density in the Cloverleaf suggested by
the HCO$^+$ and CO observations is also consistent with its location
on the HCO$^+$--far-infrared luminosity relation (Riechers et al.\
\citeyear{rie06}) within the framework of the model description of
Krumholz \& Thompson (\citeyear{kt07}). In fact, it may be the most
direct evidence that the increasing slope in dense gas-star formation
relations observed toward the most luminous, high redshift systems
(Gao et al.\ \citeyear{gao07}; Riechers et al.\ \citeyear{rie07b}) is
indeed related to an elevated median gas density relative to
lower-luminosity systems.

Besides APM\,08279+5255 ($z$=3.91), the Cloverleaf is only the second
high-$z$ galaxy in which multiple transitions of a dense gas tracer
were detected (e.g., Wagg et al.\ \citeyear{wag05}; Garcia-Burillo et
al.\ \citeyear{gar06}; Wei\ss\ et al.\ \citeyear{wei07}; Guelin et
al.\ \citeyear{gue07}; Riechers et al.\ \citeyear{rie09},
\citeyear{rie10}). Modeling of the HCN line ladder in APM\,08279+5255
suggests that the emission in high-$J$ HCN transitions is
substantially enhanced by radiative excitation through pumping of
mid-infrared ro-vibrational lines (Wei\ss\ et al.\ \citeyear{wei07};
Riechers et al.\
\citeyear{rie10}). In contrast, the HCO$^+$ excitation in the
Cloverleaf is consistent with purely collisional excitation. Given the
comparable critical densities of HCN and HCO$^+$, this suggests that
we have identified a clear difference in the dense gas excitation
conditions between these two high-$z$ systems. 

This investigation highlights the importance of studying the
excitation of dense gas tracers to understand differences in the
conditions for star formation in high redshift galaxies. Such studies
will become routine with the advent of broad instantaneous bandwidth
systems as part of future facilities such as the Atacama Large (sub-)
Millimeter Array (ALMA), which will allow to frequently cover lines of
multiple dense gas tracers as part of `standard' high-$z$ CO
observations.

\acknowledgments 
We thank the referee for helpful suggestions, Christian Henkel for the
original version of the LVG code, and Jean-Paul Kneib for a CO image
of the Cloverleaf. DR acknowledges support from from NASA through
Hubble Fellowship grant HST-HF-51235.01 awarded by STScI, operated by
AURA for NASA, under contract NAS 5-26555.  The IRAM PdBI is supported
by INSU/CNRS (France), MPG (Germany) and IGN (Spain).

\end{document}